\documentclass{aa}
\usepackage{graphics}
\usepackage[varg]{txfonts}
\usepackage{natbib}
\usepackage{hyperref}

\newcommand{\zav}[1]{\left(#1\right)}
\newcommand{\hzav}[1]{\left[#1\right]}
\newcommand{\szav}[1]{\left\{#1\right\}}

\newcommand{\kms}{\ensuremath{\text{km}\,\text{s}^{-1}}}
\newcommand{\vel}{{v}}
\newcommand{\vinfty}{\ensuremath{\vel_\infty}}
\newcommand{\vesc}{\ensuremath{\vel_\text{esc}}}
\newcommand{\ms}{\ensuremath{{M}_{\odot}}}
\newcommand{\msr}{\ensuremath{\ms\,\text{yr}^{-1}}}

\newcommand{\teff}{\ensuremath{T_\text{eff}}}

\newcommand\cc{\ensuremath{C_\text{c}}}

\begin{document}

\title{New mass-loss rates of Magellanic Cloud B supergiants from global wind
models}

\author{J.~Krti\v{c}ka\inst{1} \and J.~Kub\'at\inst{2} \and
        I.~Krti\v ckov\'a\inst{1}}

\institute{Department of Theoretical Physics and Astrophysics, Faculty of
           Science, Masaryk University, CZ-611 37 Brno,
           Czech Republic \and
           Astronomical Institute, Czech Academy of Sciences,
CZ-251 65
           Ond\v rejov, Czech Republic }

\date{Received}

\abstract{We provide global models of line-driven winds of B supergiants for
metallicities corresponding to the Large and Small Magellanic Clouds. The
velocity and density structure of the models is determined consistently from
hydrodynamical equations with radiative force derived in the comoving frame and
level populations computed from kinetic equilibrium equations. We provide a
formula expressing the predicted mass-loss rates in terms of stellar luminosity,
effective temperature, and metallicity. Predicted wind mass-loss rates decrease
with decreasing metallicity as $\dot M\sim Z^{0.60}$ and are proportional to the
stellar luminosity. The mass-loss rates increase below the region of the
bistability jump at about 20\,kK because of iron recombination. In agreement with
previous theoretical and observational studies, we find a smooth change of wind
properties in the region of the bistability jump. With decreasing metallicity,
the bistability jump becomes weaker and shifts to lower effective temperatures.
At lower metallicities above the bistability jump, our predictions provide
similar rates to those used in current evolutionary models, but our rates are
significantly lower than older predictions below the bistability jump. Our
predicted mass-loss rates agree with observational estimates derived from
H$\alpha$ line assuming that observations of stellar winds from Galaxy and the
Magellanic Clouds are uniformly affected by clumping. The models nicely
reproduce the dependence of terminal velocities on temperature derived from
ultraviolet spectroscopy.}

    \keywords{stars: winds, outflows --
              stars:   mass-loss  --
              stars:  early-type --
              supergiants --
              hydrodynamics --
              Magellanic Clouds
}

\maketitle

\section{Introduction}

The stellar winds of hot stars are accelerated by the radiative force mainly due to
line transitions of heavy elements. Thanks to their resonance character, line
transitions due to bound electrons can be much more efficient in radiative
acceleration than scattering on free electrons \citep{gamom}. However, the line
interaction is confined to a relatively narrow range of frequencies given by
line broadening. This is the place where the Doppler effect comes into play, which
enables a given line to cover a much larger portion of the stellar flux than just
the confined range of frequencies given by the line broadening \citep{milne}.

Trace elements heavier than hydrogen and helium are the most effective in absorbing stellar radiation and in accelerating the wind. Therefore, the process of
Coulomb collision between charged particles is needed to accelerate the bulk of
the wind material composed of hydrogen and helium \citep{cak76,treni,gla}.

As a result of the dependence of the radiative force on the wind chemical
composition, the amount of mass lost by the star per unit of time, that is, the wind
mass-loss rate $\dot M$, scales with metallicity
\citep{vikolamet,mcmfkont,bjorko}. In general, the winds are the strongest at
high metallicities and cease at zero metallicity \citep{bezvi}. Therefore, in addition to luminosity and effective temperature, metallicity is one of the
main parameters that determine the wind mass-loss rate.

The metallicity dependence of wind strength becomes important in galaxies that
have different star formation histories from that of our Galaxy. The Magellanic Clouds
are typical examples of such galaxies, which, thanks to their lower metallicity
\citep[e.g.,][]{kopec,ven,rolles,bourak} and relative proximity, provide an
important benchmark for the study of evolution of stars and their interaction
with the circumstellar environment. The Large and Small Magellanic Clouds (LMC
and SMC, respectively) are galaxies that gravitationally interact with our
Galaxy and the LMC is still likely completing its first orbit around the Milky Way
\citep{patel}. The LMC is massive enough to significantly perturb the orbits of the
satellites of our Galaxy \citep{bata}. The Magellanic Clouds have a complex star
formation history, with several synchronized peaks having taken place over the last few billion
years \citep{bilymracna,massamrac}. 

Mass loss influences the evolution of stars \citep{kostel} and their interaction
with the interstellar environment \citep{kobul}, which can leave impacts on future
stellar generations \citep{dorrych}. Therefore, reliable estimates of mass-loss
rates are required in order to understand the physics of stars and of the interstellar medium.
This becomes especially appealing when significantly lower wind mass-loss rates
are predicted from global wind models of massive stars
\citep{cmfkont,powrdyn,bjorko}. These models solve the radiative transfer
equation precisely in the comoving frame and allow a smooth transition from
photosphere to the wind. The suggested reduction in mass-loss rates seems to be
in line with a more precise observational analysis taking into account the
effect of wind inhomogeneities \citep{najaro,bouhil,clres2}.

Here, we provide new mass-loss rate estimates for B supergiants for metallicities
corresponding to the Magellanic Clouds. These models complement our previous
grids computed for B supergiants from our Galaxy \citep{bcmfkont} and for O
stars from the Magellanic Clouds \citep{mcmfkont} and our Galaxy
\citep{cmfkont}. Our models provide an alternative prescription to that of
\citet{bjorevol}. The mass-loss-rate prescriptions given in \citet{mcmfkont} and
\cite{bcmfkont} yield a reasonably good fit for most blue supergiant stars
analyzed by \cite{berper} and for stars from the Large Magellanic Cloud cluster R136
\citep{hezkysedi}.

\section{Wind modeling of LMC and SMC B supergiants}

We used the code METUJE, which was applied by \citet{bcmfkont} in the modeling
of Galactic B supergiants. The adopted assumptions were described by
\citet{cmfkont} in detail. The code provides spherically symmetric and
stationary models of the line-driven stellar wind by solving hydrodynamical
equations together with radiative transfer and kinetic equilibrium equations.
The equations are solved in a global manner, that is, from nearly hydrostatic
photosphere to supersonically expanding wind. The output of the model
calculations provides radial variations of structural parameters, including wind
velocity, density, temperature, and level populations. Our approach is similar
to using other codes that provide consistent hydrodynamical models of hot star winds
\citep{grahamz,powrdyn,sundyn}.

Level occupation numbers of model ions are determined from the kinetic
equilibrium equations. These equations account
for bound--bound, bound--free, and Auger transitions between individual levels.
We either adopted the models of ions from the TLUSTY model input data
\citep{ostar2003,bstar2006} or prepared them ourselves using data from the Opacity and
Iron Projects \citep{topt,zel0} and the NIST database \citep{nist}.

The derived level populations enter the absorption and emission terms in the
radiative transfer equation, which is solved in the comoving frame
\citep[CMF;][]{mikuh}. The derived radiation field is used to determine the
radiative force and radiative heating and cooling terms, which are included in
equation of motion and in energy equations. These hydrodynamical equations are
supplemented with the continuity equation and yield the radial profiles of
velocity, density, and temperature of the wind.

All equations are solved iteratively, which allows us to derive a consistent flow
structure. The models are parameterized by basic stellar parameters, that is,
the stellar effective temperature \teff, stellar mass $M$, radius $R_*$, and
surface chemical composition. The main output parameters are wind mass-loss rate
$\dot M$ and terminal wind velocity $v_\infty$.

Hot star winds show signatures of small-scale structure (clumping) in their
spectra \citep{chuchcar,clres1,floraslup}. Such inhomogeneities probably
originate from line-deshadowing instability \citep{felto,sundsim}. To
understand the effect of inhomogeneities, we calculated an additional set of models
that account for clumping \citep[as in, e.g.,][]{irchuch}. We assumed optically
thin clumping parameterized by a clumping factor
$\cc={\langle\rho^2\rangle}/{\langle\rho\rangle^2}$. Here, the angle brackets
denote the average over volume. We adopted an empirically motivated
radial-dependent clumping factor \citep{najradchuch,bouhil}:
\begin{equation}
\label{najc}
\cc(r)=C_1+(1-C_1) \, e^{-\frac{\vel(r)}{C_2}},
\end{equation}
which grows from $\cc=1$ (corresponding to smooth wind) to $C_1$ for velocities
higher than $C_2$. We selected $C_1=10$, which is roughly the value at which the
empirical H$\alpha$ mass-loss rates of O stars agree with other estimates
\citep{cmfkont} and $C_2=100\,\kms$, which is typically derived from the
observational study of \citet{bouhil}. The radial velocity variation
$\vel(r)$ in Eq.~\eqref{najc} is approximated by the fit of the wind velocity
determined for the unclumped wind ($\cc=1$) by a polynomial formula provided by
\citet{betyna}, which was modified to
\begin{equation}
\label{vrfit}
\tilde \vel (r)=\sum_i \varv_i\zav{1-\gamma\frac{R_*}{r}}^i,
\end{equation}
where $\varv_i$ and $\gamma$ are parameters of the fit.

\section{Calculated wind models}

The models were calculated for the same grid of stellar parameters as in
\citet{bcmfkont}. The parameters given in Table~\ref{bvele} cover the spectral
range of B supergiants at three different values of stellar luminosity $L$.
The adopted parameters correspond to typical values determined for B supergiants
from observations in \citet{vysbeta}. To account for a typical chemical
composition of the LMC and SMC
\citep[e.g.,][]{kopec,ven,rolles,bourak}, we calculated two sets of models with
solar chemical composition $Z_\odot$ \citep{asp09} scaled by a factor of 0.5 or
0.2 for elements heavier than helium.

The resulting wind parameters are given in Table~\ref{bvele} for these two
selected metallicities. The parameters are compared to predictions derived for
Galactic B supergiants \citep{bcmfkont} and are given for a smooth wind with
$C_1=1$ and for a clumped wind with $C_1=10$. The parameters of the fit of the
velocities in Eq.~\eqref{vrfit} ---used to calculate clumped models--- are given in
Table~\ref{mbvelech}.

\begin{table*}[t]
\caption{Stellar parameters of the model grid with derived values of the
terminal velocity $v_\infty$ (in \kms) and the mass-loss rate $\dot M$ (in \msr).}
\centering
\label{bvele}
\begin{tabular}{ccccccccccccc}
\hline
\hline
&&&\multicolumn{2}{c}{$Z=Z_\odot$}&
\multicolumn{4}{c}{$Z=0.5\,Z_\odot$}&
\multicolumn{4}{c}{$Z=0.2\,Z_\odot$}\\
&&&\multicolumn{2}{c}{$C_1=1$}&
\multicolumn{2}{c}{$C_1=1$}& \multicolumn{2}{c}{$C_1=10$}& 
\multicolumn{2}{c}{$C_1=1$}& \multicolumn{2}{c}{$C_1=10$}\\
Model &$\teff$ & $R_{*}$ &
$\vinfty$ & $\dot M$ &
$\vinfty$ & $\dot M$ &
$\vinfty$ & $\dot M$ &
$\vinfty$ & $\dot M$ &
$\vinfty$ & $\dot M$ \\
&$[\text{kK}]$ & $[{R}_{\odot}]$ \\
\hline
\multicolumn{13}{c}{$M=25\,{M}_{\odot}$, $\log(L/L_\odot)=5.28$,
$\Gamma=0.18$}\\
275-25 & 27.5 & 19.3 & 1890 & $9.1\times10^{-8}$ & 1920 & $7.7\times10^{-8}$ & 2040 & $8.4\times10^{-8}$ &1880 & $5.9\times10^{-8}$ & 2090 & $6.5\times10^{-8}$\\
250-25 & 25.0 & 23.3 & 1640 & $7.5\times10^{-8}$ & 1540 & $6.7\times10^{-8}$ & 1730 & $7.4\times10^{-8}$ &1340 & $5.5\times10^{-8}$ & 1750 & $5.5\times10^{-8}$\\
225-25 & 22.5 & 28.8 & 1130 & $7.4\times10^{-8}$ & 1060 & $6.3\times10^{-8}$ & 1300 & $7.3\times10^{-8}$ &1240 & $4.6\times10^{-8}$ & 1500 & $5.1\times10^{-8}$\\
200-25 & 20.0 & 36.4 &  760 & $7.9\times10^{-8}$ & 1520 & $4.0\times10^{-8}$ &  960 & $1.3\times10^{-7}$ &1220 & $3.9\times10^{-8}$ &  960 & $5.2\times10^{-8}$\\
175-25 & 17.5 & 47.6 &  570 & $1.4\times10^{-7}$ &  380 & $5.9\times10^{-8}$ &  820 & $1.9\times10^{-7}$ &1020 & $2.5\times10^{-8}$ &  720 & $7.5\times10^{-8}$\\
150-25 & 15.0 & 64.8 &  510 & $3.1\times10^{-7}$ &  470 & $1.5\times10^{-7}$ &  510 & $2.6\times10^{-7}$ &1050 & $1.3\times10^{-8}$ &  670 & $4.5\times10^{-8}$\\\
125-25 & 12.5 & 93.3 &  120 & $1.7\times10^{-7}$ &  130 & $1.2\times10^{-7}$ &  120 & $1.3\times10^{-7}$ & 120 & $3.5\times10^{-8}$ &  130 & $4.7\times10^{-8}$\\
100-25 & 10.0 & 146  &  480 & $8.8\times10^{-9}$ &  410 & $1.1\times10^{-8}$ &  430 & $1.1\times10^{-8}$ & 360 & $8.6\times10^{-9}$ &  430 & $9.6\times10^{-9}$\\
\hline
\multicolumn{13}{c}{$M=40\,{M}_{\odot}$, $\log(L/L_\odot)=5.66$,
$\Gamma=0.27$}\\
275-40 & 27.5 & 29.9 & 1300 & $3.4\times10^{-7}$ & 1510 & $2.6\times10^{-7}$ & 1370 & $3.1\times10^{-7}$ & 1670 & $1.4\times10^{-7}$ & 1960 & $2.0\times10^{-7}$\\
250-40 & 25.0 & 36.1 & 1600 & $2.0\times10^{-7}$ & 1320 & $1.9\times10^{-7}$ & 1610 & $2.1\times10^{-7}$ & 1170 & $1.5\times10^{-7}$ & 1500 & $1.6\times10^{-7}$\\
225-40 & 22.5 & 44.6 & 1160 & $2.0\times10^{-7}$ &  880 & $2.0\times10^{-7}$ &  820 & $2.7\times10^{-7}$ &  860 & $1.6\times10^{-7}$ & 1220 & $1.7\times10^{-7}$\\
200-40 & 20.0 & 56.4 &  700 & $2.6\times10^{-7}$ &  670 & $1.9\times10^{-7}$ & 1100 & $4.6\times10^{-7}$ &  930 & $1.1\times10^{-7}$ &  890 & $1.9\times10^{-7}$\\
175-40 & 17.5 & 73.7 &  630 & $6.3\times10^{-7}$ &  580 & $3.2\times10^{-7}$ &  520 & $8.6\times10^{-7}$ &  530 & $1.0\times10^{-7}$ &  780 & $2.8\times10^{-7}$\\
150-40 & 15.0 & 100  &  110 & $1.5\times10^{-6}$ &   90 & $7.3\times10^{-7}$ &  110 & $1.2\times10^{-6}$ &   60 & $1.5\times10^{-7}$ &   80 & $5.0\times10^{-7}$\\
125-40 & 12.5 & 145  &   80 & $6.6\times10^{-7}$ &  100 & $4.3\times10^{-7}$ &  100 & $4.7\times10^{-7}$ &   70 & $1.7\times10^{-7}$ &  110 & $1.9\times10^{-7}$\\
100-40 & 10.0 & 226  &  410 & $2.9\times10^{-8}$ &  410 & $3.3\times10^{-8}$ &  440 & $3.4\times10^{-8}$ &  400 & $2.7\times10^{-8}$ &  410 & $2.7\times10^{-8}$\\
\hline
\multicolumn{13}{c}{$M=60\,{M}_{\odot}$, $\log(L/L_\odot)=5.88$,
$\Gamma=0.30$}\\
275-60 & 27.5 & 38.5 & 1240 & $6.1\times10^{-7}$ & 1680 & $4.1\times10^{-7}$ & 1520 & $4.9\times10^{-7}$ & 1910 & $3.0\times10^{-7}$ & 1930 & $3.6\times10^{-7}$\\
250-60 & 25.0 & 46.5 & 1850 & $3.2\times10^{-7}$ & 1780 & $2.8\times10^{-7}$ & 1720 & $3.4\times10^{-7}$ & 1520 & $2.3\times10^{-7}$ & 1860 & $2.5\times10^{-7}$\\
225-60 & 22.5 & 57.5 & 1000 & $3.4\times10^{-7}$ &  880 & $3.1\times10^{-7}$ & 1310 & $4.6\times10^{-7}$ &  940 & $2.6\times10^{-7}$ & 1250 & $2.8\times10^{-7}$\\
200-60 & 20.0 & 72.7 &  760 & $4.3\times10^{-7}$ &  650 & $3.2\times10^{-7}$ & 1230 & $8.8\times10^{-7}$ & 1170 & $1.7\times10^{-7}$ & 1080 & $2.7\times10^{-7}$\\
175-60 & 17.5 & 95.0 &  820 & $1.2\times10^{-6}$ &  650 & $5.2\times10^{-7}$ &  800 & $1.4\times10^{-6}$ & 1570 & $1.0\times10^{-7}$ &  940 & $2.4\times10^{-7}$\\
150-60 & 15.0 & 129  &  490 & $2.1\times10^{-6}$ &  540 & $1.1\times10^{-6}$ &  530 & $1.6\times10^{-6}$ &  460 & $3.9\times10^{-7}$ &  580 & $4.9\times10^{-7}$\\
125-60 & 12.5 & 186  &  110 & $1.3\times10^{-6}$ &  150 & $8.1\times10^{-7}$ &  140 & $8.9\times10^{-7}$ &  130 & $3.1\times10^{-7}$ &  140 & $3.8\times10^{-7}$\\\
100-60 & 10.0 & 291  &  120 & $6.8\times10^{-8}$ &  120 & $6.0\times10^{-8}$ &  120 & $6.2\times10^{-8}$ &  130 & $4.8\times10^{-8}$ &  110 & $4.5\times10^{-8}$\\
\hline
\end{tabular}
\tablefoot{$\Gamma$ is the Eddington parameter for electron Thomson scattering
opacity.}
\end{table*}

\begin{table*}[t]
\caption{Parameters of the fit of the wind radial velocity of unclumped model
(Eq.~\eqref{vrfit}).}
\centering
\label{mbvelech}
\begin{tabular}{crrrrrrrr}
\hline
\hline
      & \multicolumn{4}{c}{$0.5Z_\odot$} & \multicolumn{4}{c}{$0.2Z_\odot$}\\
Model & $\varv_1$ & $\varv_2$ & $\varv_3$ & $\gamma$ & $\varv_1$ & $\varv_2$ &
 $\varv_3$ & $\gamma$\\
      & \multicolumn{3}{c}{[\kms]} & & \multicolumn{3}{c}{[\kms]}\\
\hline
275-25 & 3720 & $-$1652 &       & 1.042 & 3508 & $-$1549 &       &    1.044\\
250-25 & 4694 & $-$3631 &       & 1.045 & 3812 & $-$2771 &       &    1.044\\
225-25 & 3608 & $-$3161 &       & 1.059 & 1853 &  $-$597 &       &    1.040\\
200-25 & 2105 &  $-$493 &       & 1.043 & 1143 &     244 &       &    1.039\\
175-25 &  836 &  $-$293 &       & 1.064 & 2014 &  $-$781 &       &    1.072\\
150-25 &  646 &   $-$92 &       & 1.069 & 2603 & $-$1620 &       &    1.101\\
125-25 &  567 & $-$1117 &   797 & 1.078 &  437 &  $-$775 &  500  &    1.076\\
100-25 &  790 &  $-$346 &       & 1.129 &  440 &   $-$35 &       &    1.111\\
\hline                                
275-40 & 2380 &  $-$620 &       & 1.056 & 3954 & $-$2269 &       &    1.069\\
250-40 & 3519 & $-$2256 &       & 1.058 & 3836 & $-$3238 &       &    1.057\\
225-40 & 2663 & $-$2021 &       & 1.065 & 2496 & $-$1907 &       &    1.059\\
200-40 & 1659 & $-$1043 &       & 1.070 & 1527 &  $-$419 &       &    1.064\\
175-40 &  790 &   $-$91 &       & 1.074 &  848 &   $-$48 &       &    1.076\\
150-40 &  724 & $-$1891 &  1532 & 1.093 &  630 & $-$2454 & 3276  &    1.096\\
125-40 &  625 & $-$1635 &  1519 & 1.103 &  779 & $-$3351 & 5345  &    1.115\\
100-40 &  763 &  $-$321 &       & 1.155 &  537 &  $-$106 &       &    1.149\\
\hline                                
275-60 & 2629 &  $-$753 &       & 1.057 & 2723 &  $-$568 &       &    1.059\\
250-60 & 4737 & $-$3096 &       & 1.066 & 4854 & $-$3956 &       &    1.062\\
225-60 & 2731 & $-$2155 &       & 1.062 & 2465 & $-$1675 &       &    1.052\\
200-60 & 1770 & $-$1200 &       & 1.068 & 1392 &      94 &       &    1.061\\
175-60 &  851 &   $-$98 &       & 1.069 & 2471 &  $-$688 &       &    1.092\\
150-60 &  808 &  $-$168 &       & 1.075 &  635 &   $-$78 &       &    1.075\\
125-60 &  539 &  $-$939 &   598 & 1.084 &  502 &  $-$932 &  620  &    1.100\\
100-60 &  619 & $-$1363 &  1081 & 1.126 &  505 &  $-$902 &  603  &    1.124\\
\hline
\end{tabular}
\end{table*}

\begin{figure}
\includegraphics[width=0.5\textwidth]{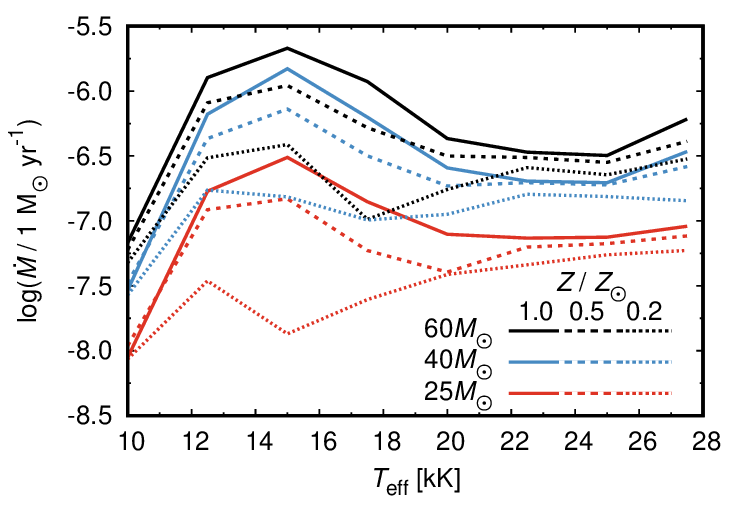}
\caption{Predicted mass-loss rate as a function of effective temperature.
Line color denotes stellar mass (or luminosity) and line type
encodes metallicity, with solid lines corresponding to $Z_\odot$, dashed lines to
$0.5\,Z_\odot$, and dotted lines to $0.2\,Z_\odot$.}
\label{mbdmdttep}
\end{figure}

From Fig.~\ref{mbdmdttep} and Table \ref{bvele}, it follows that the mass-loss
rate increases with increasing luminosity and metallicity. On average, the
mass-loss rates of B supergiants
increase with metallicity as $\dot M\sim Z^{0.60}$.
%fitovani log(Z):log(dMdt/dMdt_odot) funkci f(x)=x*a
The temperature variations are more complex. The mass-loss rate slightly
decreases with decreasing temperature down to about 22\,kK because of the declining flux
in the ultraviolet domain, where the lines accelerating the wind mostly appear.
Below this temperature, the mass-loss rates start to gradually increase by a factor of a few.
This rise is referred to as a bistability jump \citep{mnichov8}, and is caused
by the recombination of iron from \ion{Fe}{iv} to \ion{Fe}{iii}
\citep{vikolabis,vinbisja,bcmfkont}. The change in the efficiency of wind
acceleration is connected to the fact that ions with lower charge have
typically more lines at longer wavelengths close to the stellar flux maximum.

\begin{figure}[t]
\centering
\resizebox{\hsize}{!}{\includegraphics{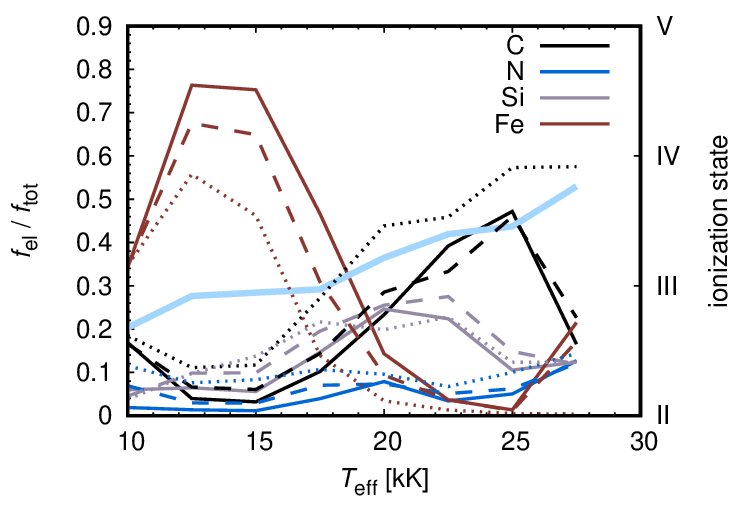}}
\caption{Relative contribution of selected elements to the line radiative force
as a function of the effective temperature of a model star, plotted at the wind
critical point for models with $M=40\,M_\odot$. Different line styles denote
metallicity, with the solid line corresponding to $Z_\odot$, dashed line to
$0.5Z_\odot$, and dotted line to $0.2Z_\odot$. The thick blue line (with
corresponding axis on the right) plots the mean ionization state driving the
wind for metallicity $0.5Z_\odot$. This is defined as $\sum f_iz_i/(\sum f_i)$,
where $f_i$ is the contribution of ion $i$ with charge $z_i$ to the line
radiative force. For simplicity, the values were obtained within the Sobolev
approximation.}
\label{mbvsil}
\end{figure}

These effects are also demonstrated in Fig.~\ref{mbvsil}, which plots the
contribution of selected elements to the radiative force as a function of
effective temperature. In comparison with Fig.\,2 in \citet{bcmfkont}, we do
not plot the elements with a relatively low contribution (Al, P, and S). For the
hottest stars, the radiative force is dominated by elements lighter than iron.
With decreasing temperature the wind recombines and the radiative driving
becomes dominated by iron for $\teff<15\,\text{kK}$. However, due to lower iron
abundance, with decreasing metallicity the iron lines become optically thin and
contribute to the radiative force to a lesser extent. As a result, the
bistability jump weakens with lower metallicity \citep[as found by][]{vikolamet}
and becomes apparent only when \ion{Fe}{iii} dominates for
$\teff<15\,\text{kK}$.

\begin{table*}[t]
\caption{Parameters of the fit of the mass-loss rate in Eq.~\eqref{dmdtobz}.}
\label{fit}
\begin{tabular}{*{10}{c}}
\hline
\hline
$a$ & $b$ & $c$ & $T_1$ & $T_2$ & $\Delta T_1$ & $\Delta T_2$ & $\alpha$ &
$\beta$ & $\delta$ \\
&&& [kK] & [kK]& [kK]& [kK]\\
\hline
$-$13.82 & 1.52 & 3.84 & 14.16 & 37.90 & 3.58 & 56.5 & 0.358 & $-$0.11 & 0.73 \\
\hline
\end{tabular}
\centering
\end{table*}

The predicted mass-loss rates of OB supergiants and main sequence O stars, and
giants from the Galaxy and the Magellanic Clouds without clumping
\citep{cmfkont,mcmfkont,bcmfkont}, can be fitted via 
\begin{multline}
\label{dmdtobz}
\log\zav{\frac{\dot M}{1\, \msr }}= a + \alpha\log\zav{\frac{Z}{Z_\odot}}+
\hzav{b+\beta\log\zav{\frac{Z}{Z_\odot}}} \log\zav{\frac{L}{10^6L_\odot}} \\
-a
\log\szav{\hzav{1+\delta\log\zav{\frac{Z}{Z_\odot}}}\exp\hzav{-\frac{\zav{\teff-T_1}^2}{\Delta
T_1^2}}\right.\\\left.+
c \exp\hzav{-\frac{\zav{\teff-T_2}^2}{\Delta T_2^2}}}.
\end{multline}
Here, $\log$ denotes logarithm with base 10 and $\exp$ denotes the natural exponential.
Derived parameters of the fit are given in Table~\ref{fit}. The fit approximates
the predicted mass-loss rates with a typical error of about 20 \%. The error is
slightly higher in the B supergiant domain. The fit Eq.~\eqref{dmdtobz} uses a
Gaussian function to depict the temperature modulation of the mass-loss rates,
and therefore it cannot be applied outside the fitted $\teff$ domain, which is 10 --
45\,kK. Application of Eq.~\eqref{dmdtobz} for luminosities and metallicities
significantly outside the domain for which we obtained the fit, that is,
$\log\zav{{L}/{L_\odot}}= 5.28-5.88$ and $Z/Z_\odot=0.2-1.0$, could also lead to
spurious results.

In addition to the discussed dependence of the mass-loss rate on the stellar
luminosity, effective temperature, and metallicity, Eq.~\eqref{dmdtobz}
introduces scaling of luminosity and temperature dependencies with the
metallicity via parameters $\beta$ and $\delta$. The temperature dependence is
modified only below the bistability jump thanks to the absence of metallicity in the
very last term of Eq.~\eqref{dmdtobz}. These modifications lead to a steeper
dependence of the mass-loss rate on luminosity for lower metallicities and to a
weaker bistability jump at low metallicities. The modification of the luminosity
dependence of the mass-loss rate with metallicity is very similar to that
introduced by \citet{mcmfkont}, who interpreted the modification in terms of
weaker blocking of far-UV flux at low metallicity and by a variation of the
slope of the line-strength distribution function \citep{pusle} with metallicity.
The weakening of the bistability jump at low metallicities is connected to the
lower contribution of iron to the radiative force for lower metallicities
(Fig.~\ref{mbvsil}, also \citealt{vikolamet}). \citet{bjorevol} introduced
an additional modification of temperature variations with metallicity that can
mimic luminosity variations in Eq.~\eqref{dmdtobz}, albeit with stronger effect.
Analogous but steeper metallicity variations were found by \citet{alexvyvoj}.
\citet{visamet} predict the metallicity dependence of mass-loss rates as $\dot
M\sim Z^{0.42}$ above the bistability jump and $\dot M\sim Z^{0.85}$ below the
jump. For a typical luminosity in their work, $\log(L/L_\odot)=5.5$,
Eq.~\eqref{dmdtobz} predicts very similar scaling $\dot M\sim Z^{0.41}$ above
the jump and a dependence below the jump that is stronger by a factor of a few. 

As a result of clumping, the matter density in clumps becomes higher than the
density in a smooth wind. For optically thin clumping, this leads to stronger
recombination \citep{hamko,bourak,martclump,pulchuch}. Because ions with lower
charge are typically able to drive wind more efficiently, with stronger clumping
the mass-loss rate increases (Table~\ref{bvele}). Typically, the increase is
relatively small, but this effect intensifies in the region of the bistability
jump. As a result of clumping, the bistability jump shifts towards higher
effective temperatures. This leads to a more significant increase in the
mass-loss rate by a factor of up to three, which corresponds to a proportionality of
$\dot M\sim\cc^{0.5}$ there. Outside the region of the bistability jump (for
$\teff\geq22.5\,$kK or $\teff\leq12.5\,$kK), the increase in the mass-loss rate
is lower, of namely about $1.14,$ corresponding to $\dot M\sim\cc^{0.06}$.

Wind theory predicts that the terminal velocities scale mostly with surface
escape speed \citep{cak}. In our sample of stars with constant luminosities,
radius increases with decreasing effective temperature, and therefore the scaling
of the terminal velocities with escape speed leads to a decrease in terminal
velocity with temperature (Table~\ref{bvele}). The terminal velocities
additionally significantly decrease in the region of the bistability jump
\citep{vinbisja,bcmfkont}, which can be interpreted in terms of a varying line-strength distribution function \citep{pusle,bcmfkont}. The terminal velocities
increase again around $\teff\approx10\,$kK as a result of the lower importance of
iron for line driving. We did not find any strong dependence of terminal
velocity on metallicity, except the slight shift of the bistability jump towards
lower temperatures with decreasing metallicity.

\section{Tests against observations and comparison with independent theoretical
models}

\begin{figure}
\includegraphics[width=0.5\textwidth]{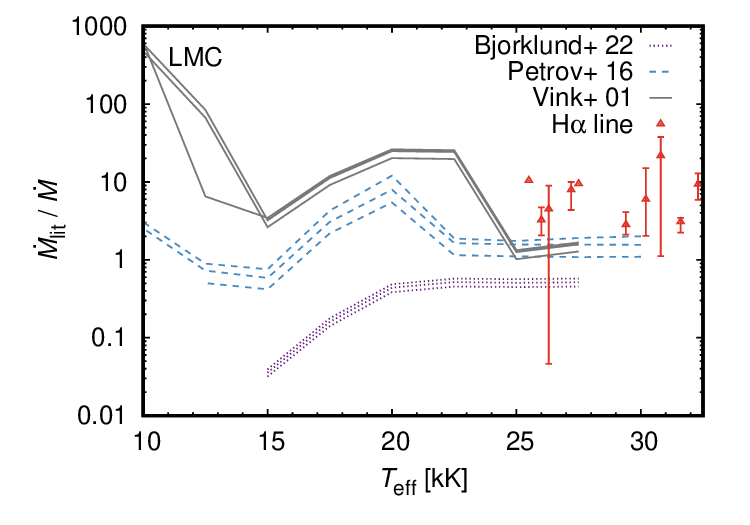}
\includegraphics[width=0.5\textwidth]{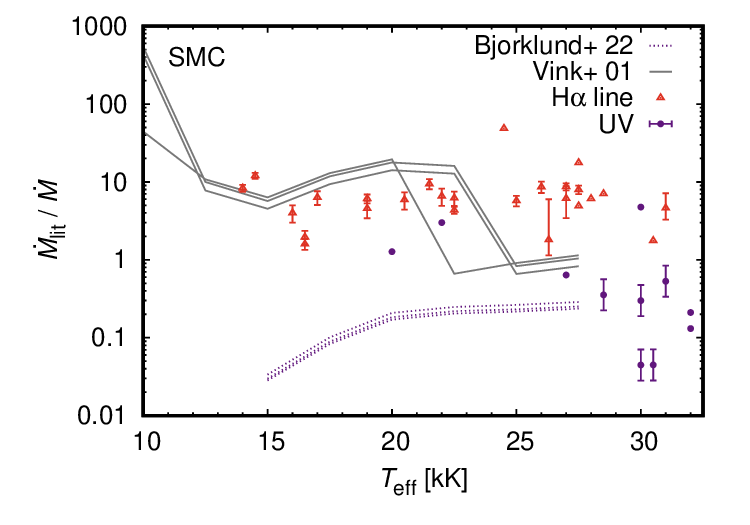}

\caption{Ratio of mass-loss rates derived from the literature and mass-loss rates
predicted using Eq.~\eqref{dmdtobz} plotted as a function of effective
temperature for stars from the LMC ({\em Upper panel}) and SMC ({\em Lower
panel}). The observational values include mass-loss rates determined from the
Balmer lines \citep[mostly H$\alpha$ line, red
triangles,][]{bezchuch,trundlebsmc1,trundlebsmc2,mokm,mokv}, and from UV wind
lines corrected for clumping \citep[violet dots,][]{ramakridlo,boumal}.
Overplotted are theoretical predictions evaluated for stellar parameters from
our sample \citep{vikolamet,bjorevol} (solid gray and dotted purple lines,
respectively) and for parameters of \citet{petcmfgen} (dashed blue lines).}
\label{dmdtpodil}
\end{figure}

While the wind mass-loss rates can be determined from manifold observables
spanning almost the entire electromagnetic domain in our Galaxy, the choice is
significantly constrained in nearby galaxies. Typically, only mass-loss rates
from optical hydrogen emission lines are available, supplemented in rare cases
with mass-loss rates from UV wind lines. This significantly complicates
any comparison with theoretical predictions because of the influence of clumping. While
estimates from UV regions can be independently corrected for clumping
\citep{ramakridlo,boumal}, the estimates based on optical data are overestimated
by a factor of a few, which depends on the strength of clumping \citep [expressed by
\cc,][] {pulvina}. Moreover, while observational studies typically account for
optically thin clumping at most, allowing for optically thick clumps provides
results that are more robust \citep{clres2}.

This is reflected in the comparison of theoretical predictions with observational
data in Fig.~\ref{dmdtpodil}. Here, we plot the ratio of mass-loss rates
determined from observations to those predicted for individual stars.
For mass-loss rates derived from optical emission lines, the ratio is always
higher than one, which indicates that the theory does not contradict
observations. Quantitatively, the mass-loss rates determined from optical
emission lines are higher than
predictions by factors of, on average, 6.4 and 6.0 for the LMC and SMC, respectively. Considering the proportionality
of predicted mass-loss rates, $\dot M\sim\cc^{0.06}$ , and the dependence of
clumping corrections via $\dot M\sim\cc^{-0.5}$ \citep[e.g.,][]{pulvina}, the
theory agrees with observations for $\cc=28$ for the LMC and for $\cc=25$ for
the LMC. These are slightly higher values than those obtained from observational
studies of SMC stars \citep{boumal}. On the other hand, these values are close
to $\cc=24,$ which is the value derived for Galactic supergiants \citep{bcmfkont}, and agree with
theory and observations showing that the (optically thin) clumping factor is
almost independent of metallicity \citep{hezkysedi,drimetal}.

The theoretical predictions of mass-loss rates agree better with estimates from
UV lines (and H$\alpha$ in some cases) corrected for optically thin
clumping. On average, the theoretical predictions are by a factor of 2.5 higher
than values from observations. However, this comparison is perhaps influenced by
the so-called "weak wind problem", which relates to the weak wind lines in the spectra
of late O stars \citep{bourak,martin}. This effect is considered to be caused by
slow cooling of wind material in post-shock regions
\citep{cobecru,nlteiii,predbehli}. Moreover, some of the stars analyzed by
\citet{ramakridlo} are main sequence stars, for which we have to extrapolate in
Eq.~\eqref{dmdtobz}.

If the effect of clumping on observations is neglected, predictions of
\citet{vikolamet} underestimate the mass-loss rates for early B SMC supergiants
and overestimate them for mid B supergiants \citep{trundlebsmc1}. This can be
seen in Fig.~\ref{dmdtpodil}, where we also plot the ratio of \citet{vikolamet}
predictions to our predictions for stars from our sample (Table~\ref{bvele}).
The \citet{vikolamet} predictions are plotted for a terminal-to-escape-speed ratio
of 2.6 above the bistability jump and 1.3 below the jump, respectively. Above
the jump and for higher $\vinfty/\vesc$ ratios and low metallicities, these
predictions reasonably agree with ours, although there is significant
disagreement for higher metallicities or higher effective temperatures
\citep{cmfkont,bcmfkont}. This leads us to the conclusion that the differences
between the rates of \citet{vikolamet} and ours are caused mostly by the treatment of
iron lines, which do not significantly contribute to the line driving for B
supergiants at low metallicities and above the bistability jump
(Fig.~\ref{mbvsil}). However, the bistability jump predicted by
\citet{vikolamet} is stronger than ours, resulting in greater disagreement in
predicted rates below about 22.5\,kK. The very large disagreement between our rates
and the predictions of \citet{vikolamet} around $\teff\approx10\,$kK
(Fig.~\ref{dmdtpodil}) is connected to the appearance of the second bistability jump, which is
due to recombination of \ion{Fe}{iii}. This is the main reason why the values of
mass-loss rates differ so markedly at the lower $\teff$ boundary of our grid. In
the CMFGEN \citep{hilmi} models provided by \citet{petcmfgen}, the recombination
appears at lower temperatures, around $\teff\approx9\,$kK. Also, in our models,
\ion{Fe}{iii} dominates at the critical point, where the mass-loss rate is
determined.

While our mass-loss rate predictions agree  reasonably well with the theoretical
calculations of \citet{bjorevol} for solar metallicities, they are by a factor
of a few higher at lower metallicities (Fig.~\ref{dmdtpodil}). The agreement is
even worse in the region below the temperature of the bistability jump. The
bistability jump is missing in \citet{bjorevol}. Except in the region of the
bistability jump around 17.5\,kK, our predictions  agree nicely with the calculations
of \citet{petcmfgen}, which are based on the CMFGEN code.

\begin{figure}
\includegraphics[width=0.5\textwidth]{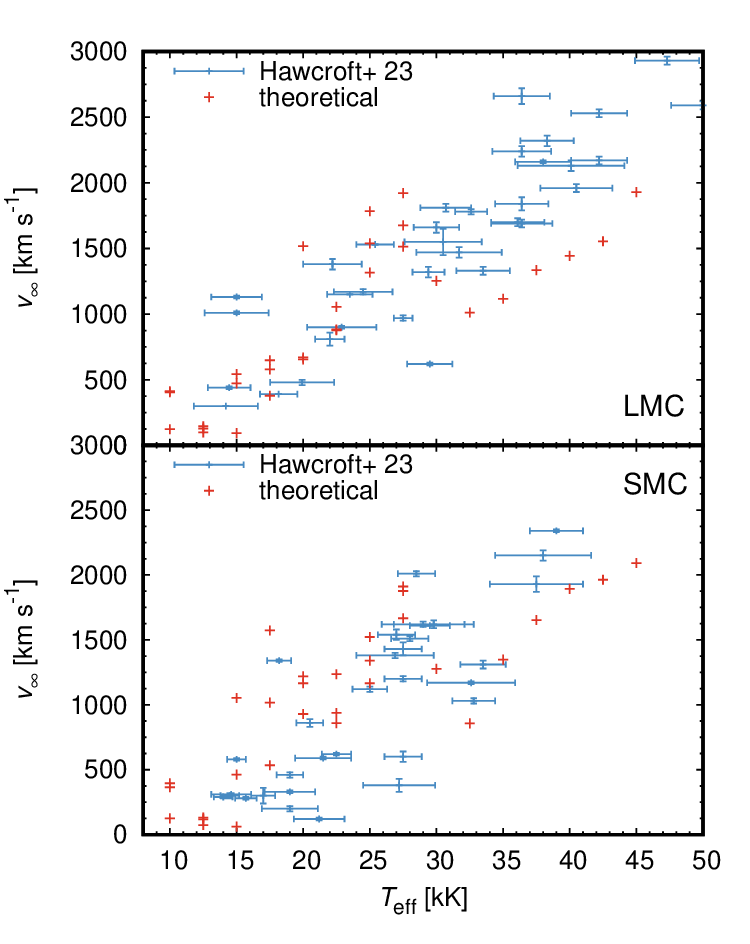}
\caption{Terminal velocities of OB supergiants from the LMC ({\em upper panel})
and SMC ({\em lower panel}) as a function of temperature \citep{snadtostihnou}
in comparison with our theoretical values.}
\label{vnektep}
\end{figure}

Observational analysis of terminal velocities provides another test of stellar
wind theory. Wind terminal velocity is typically studied in terms of the
temperature dependence of the terminal-velocity-to-escape-speed ratio
\citep{lsl,vysbeta}. However, this relationship shows relatively large scatter.
Terminal velocities can also be conveniently tested in terms of their dependence
on the effective temperature \citep{snadtostihnou}, which also reflects the
evolution of stars. We used a large sample of terminal velocity measurements of SMC
and LMC OB supergiants obtained from the ULLYSES project \citep {ulisne,xshootu}
by \citet{snadtostihnou}. These values are compared with predictions obtained
here and by \citet{bcmfkont}.

The plot (Fig.~\ref{vnektep}) shows an increase in terminal velocity with
temperature, which stems from the scaling of terminal velocity with
escape speed and from the decreasing escape speed in the course of stellar
evolution. These trends are nicely reproduced by our models, in which the
relationship between the stellar effective temperature, mass, and luminosity
reflects the evolution of stars.

The models can also be tested against observations of stars residing in
galaxies from the Local Group. \citet{sochar} derived wind parameters for two
early B-type supergiants with $Z=1.0$ and $Z=0.3$ in the Sculptor galaxy
\object{NGC 300} from optical spectroscopy. The derived mass-loss rates are higher than those predicted using Eq.~\eqref{dmdtobz} 
by a
factor of about 10. This factor
corresponds to the level of clumping found in other B supergiants. On the other
hand, the terminal velocities determined from observations nicely correspond to theoretical predictions. A similar discrepancy of a factor of about 5 in mass-loss rates
can be found from optical analysis of \object{M33} supergiants \citep [with
$Z=0.3-1.0$] {trojuhelnik}, again neglecting the influence of clumping.

\begin{figure*}
\includegraphics[width=0.5\textwidth]{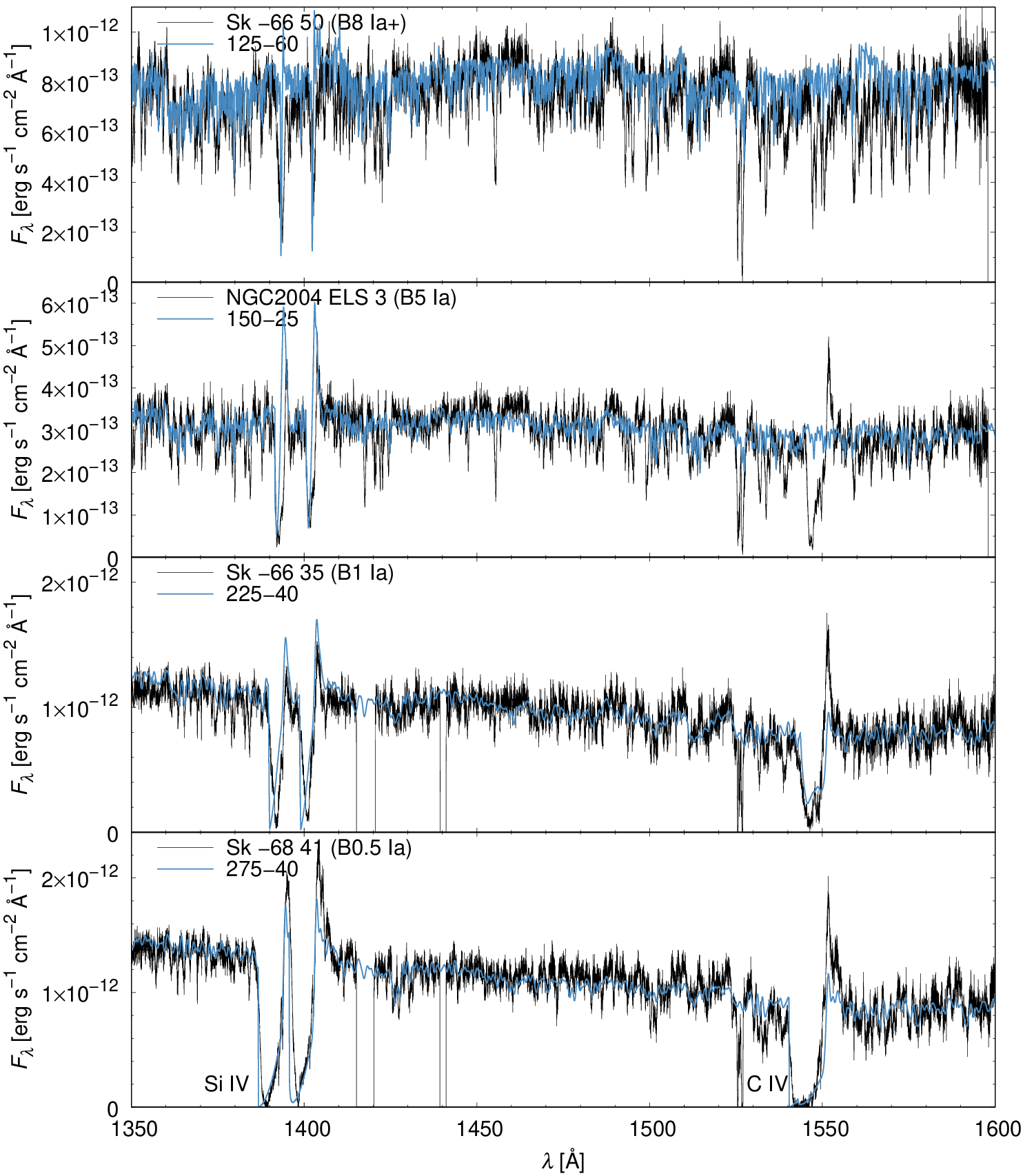}
\includegraphics[width=0.5\textwidth]{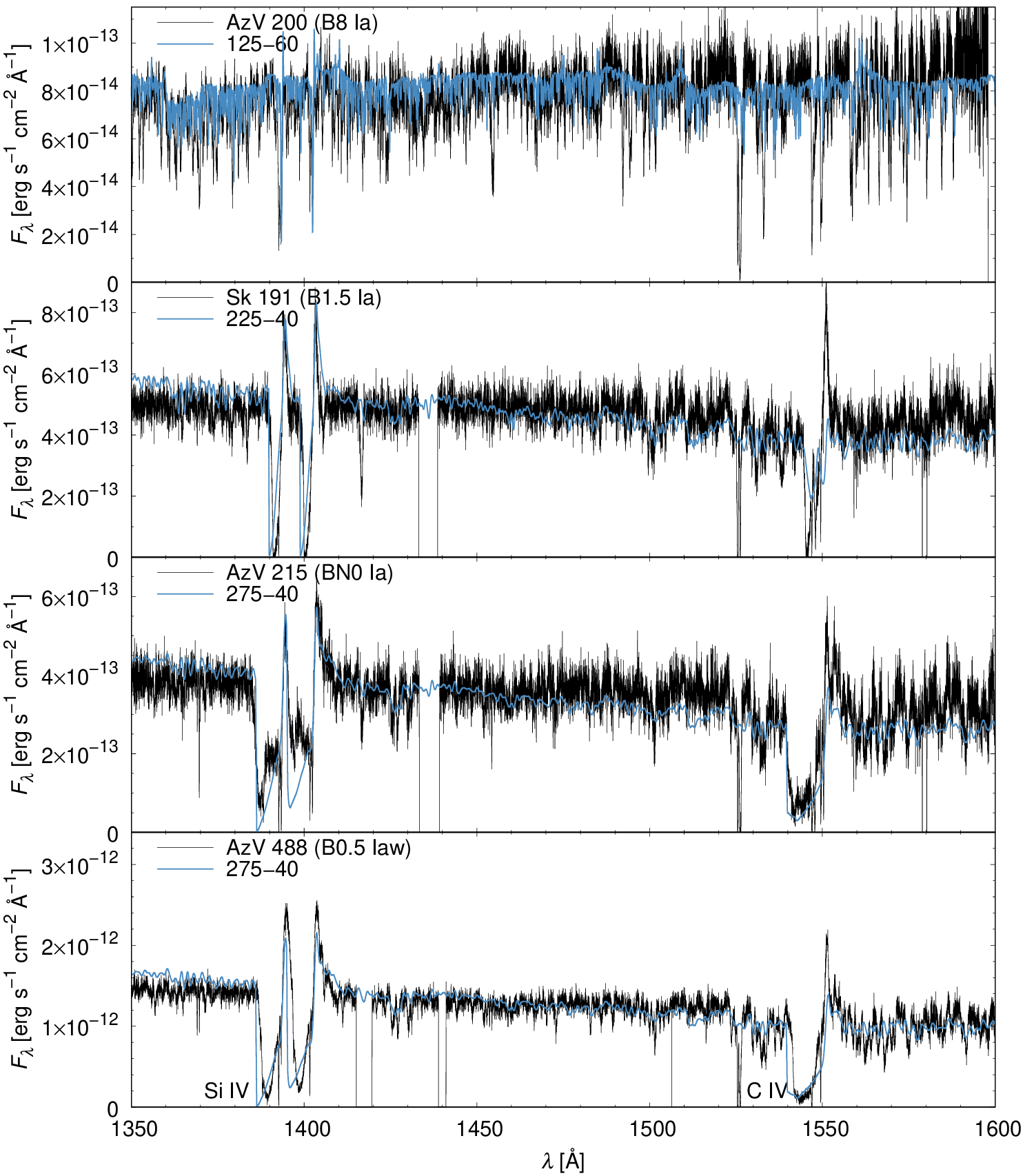}
\caption{Comparison of the spectra of selected B supergiants (black lines) with the
emergent spectra of model stars (blue lines) from the grid for the LMC ({\em
left panel}) and SMC ({\em right panel}).}
\label{spektrabvelem}
\end{figure*}

To roughly check our models with observations, in Fig.~\ref{spektrabvelem} we
provide the spectra of selected B supergiants from the LMC and SMC in the region of
strong wind lines observed by HST via the ULLYSES program \citep{ulisne} and
downloaded from the publicly available MAST
archive\footnote{https://mast.stsci.edu/search/ui/\#/ullyses}. We chose stars
that have parameters listed in \cite{snadtostihnou} that are close in value to their counterparts in the
calculated models from our grid (Table~\ref{bvele}). This series of spectra show
strengthening of wind lines towards higher effective temperatures, which is
nicely reproduced by the models. Some features are not fully reproduced because of missing physics, such as the neglect of small-scale structures
\citep{chuchcar,clres1} or X-ray emission \citep{lojza,berper}, but the overall
agreement is reasonable.

\section{Discussion}

The present paper provides further support for the downward revision of mass-loss
rate predictions in hot massive stars \citep{cmfkont,sundyn,alexvyvoj}. Together
with observational indications of lower mass-loss rates in massive stars
\citep[e.g.,][]{clres2,kobul,hezkysedi}, this challenges our current
understanding of massive star evolution. Although lower mass-loss rates might be
favorable for the solution of particular problems, such as the overly heavy
masses of merging black holes \citep{jiniabbotti,grohhmotcd}, they may give rise
to other discrepancies.

On the other hand, line-driven winds might not form a dominant constituent of
the total stellar mass loss. For instance, evolutionary models of a
solar-metallicity star with an initial mass of $60\,M_\odot$ predict that more
than half of the stellar mass is lost before the star enters the Wolf-Rayet
stage \citep{gromek}. However, only about one-third of the mass loss happens
during OB phases, the rest appears during the luminous blue variable (LBV)
stage. Evolved stars, such as LBVs, may lose mass by other mechanisms
\citep{sangras,owoex}.  In a similar way, the red supergiant phase may be
important for total mass loss in stars with lower initial mass \citep{renzovit}.

There are other observables that can be used to test the influence of mass loss
on stellar evolution; for instance, the distribution of observed rotational
velocities and the corresponding angular momentum loss \citep{kostel}. Mass loss
also modifies stellar radii and luminosities \citep{alexvyvoj}. The strength of
interaction between massive stars and surrounding circumstellar environment
depends on the wind mass-loss rate, but numerical simulations show that the
effects due to winds only dominate over the influence of ionizing flux in very
dense clouds close to the stars \citep{dalevit}. Weaker line driving leads to
weaker ablation of circumstellar discs \citep{keeobrusoebe}, which could be one
of the effects behind the appearance of relatively early Oe stars in the SMC
\citep{zlatymarx}.

\section{Conclusions}

We provide global hydrodynamical models of line-driven winds in B supergiants
from the Large and Small Magellanic Clouds. The velocity and density structure
of the models is determined using radiative force calculated from the solution
of the radiative transfer equation in the comoving frame using level populations
derived from kinetic equilibrium equations. The models are calculated in a
global manner, that is, we solve structural equations from nearly hydrostatic
atmosphere to supersonically expanding radiatively driven wind. The models are
parameterized by basic stellar parameters, which are effective temperature,
radius, mass, and metallicity.

Predicted wind mass-loss rates scale with metallicity on average as $\dot M\sim
Z^{0.60}$ and are proportional to the stellar luminosity. Mass-loss rates
are sensitive to stellar effective temperature; they significantly increase
in the region of the so-called bistability jump at about 20\,kK. However, the
variations at this temperature do not resemble a steep jump, but are rather
continuous and appear due to recombination of \ion{Fe}{iv} to \ion{Fe}{iii}.
Moreover, the strength of the jump weakens towards lower metallicities.

We provide a formula expressing predicted mass-loss rates in terms of basic
stellar parameters. At metallicities corresponding to those found in the Magellanic Clouds and
for temperatures above the bistability jump, this formula provides similar rates
to those used in current evolutionary models. However, our predictions are
significantly lower than those of \citet{vikolamet} below the bistability
jump. We compare our predictions with observational estimates derived from the
H$\alpha$ line. Above the bistability jump, H$\alpha$ mass-loss rate estimates
for stars  from the Galaxy and Magellanic Clouds agree with theoretical
predictions assuming a single clumping factor of about 25. Our models nicely
reproduce the observational dependence of terminal velocity on temperature
derived from UV spectroscopy and provide spectra that reasonably agree
with observations.

\begin{acknowledgements}
Computational resources were provided by the e-INFRA CZ project (ID:90254),
supported by the Ministry of Education, Youth and Sports of the Czech Republic.
The Astronomical
Institute Ond\v{r}ejov is supported by a project RVO:67985815 of the Academy of
Sciences of the Czech Republic.
\end{acknowledgements}

\bibliographystyle{aa}
\bibliography{papers}

\end{document}